\documentclass[11pt,twoside]{atmp}

\newtheorem{Definicion}{Definition}
\newtheorem{Ejemplo}{Example}

\newtheorem{Lema}{Lemma}
\newtheorem{Teorema}{Theorem}

\usepackage{amsmath,amssymb}

\usepackage[all]{xy}
\usepackage{color}

\begin{document}

\title{The Feynman integral as a limit of complex measures}

\author[Jos\'e L. Mart\'\i nez-Morales]{Jos\'e L. Mart\'\i nez-Morales}

\address{Instituto de Matem\'aticas, Universidad Nacional Aut\'onoma de M\'exico, \\
A.P. 273, Admon. de correos \#3\\
Cuernavaca, Morelos 62251, MEXICO\\
martinez@matcuer.unam.mx}  
\addressemail{martinez@matcuer.unam.mx}

\begin{abstract}
The fundamental solution of the Schr\"odinger equation for a free particle is a distribution. This distribution can be approximated by a sequence of smooth functions. It is defined for each one of these functions, a complex measure on the space of paths. For certain test functions, the limit of the integrals of a test function with respect to the complex measures, exists. We define the Feynman integral of one such function by this limit.
\end{abstract}

\maketitle
\section{Introduction}
The goal of the present article is to provide a mathematically sound definition of the standard Feynman path integral representation of the unitary evolution operator $e\sp{-it\hbar H}$ in the quantum mechanics which is generated by a Hamiltonian $H$=$H$* with a potential. That is, the distribution kernel of $e\sp{-it\hbar H}$ is expressed in terms of a path integral weighted with exp[$iS$($x$($\cdot$))], where $S$ is the classical action corresponding to the Hamiltonian $H$.

Although this is an old subject, we recommend to look, for example, at the references \cite{Albeverio}, \cite{Fujiwara} and \cite{Chung}. Each of them represents a distinct line of thought about the field. The first reference, in particular, gives a complete list of references.

While an analogous idea leading to path integral representations, called Feynman-Kac formulas, of the corresponding semi-group $e\sp{-\beta H}$ is firmly anchored in the theory of stochastic processes, the construction of the above Feynman path integral for $e\sp{-it\hbar H}$ as an oscillatory integral is notoriously difficult. There are various No-go theorems limiting the possibilities to begin with.

These mathematical difficulties are largely ignored by the theoretical physics community. For instance, the quantization of field theories, notably gauge theories, is entirely based on the Feynman path integral, as a formal procedure, despite its lacking of mathematical foundation.

So, for these reasons, mathematical progress towards the construction of the Feynman path integral, even if very small, deserves appreciation.

We will present the mathematical difficulties and the current developments of the Feynman path integral.
\subsection{Various known approaches to the rigorous construction of the Feynman path integral representation to the solution of the Schr\"odinger equation}
The Feynman path integral is known to be a powerful tool in different domains of
physics. At the same time, the mathematical theory underlying lots of (often formal)
physical calculations is far from being complete. Various known approaches to the rigorous construction of the Feynman path integral representation to the solution of the Schr\"odinger equation
\begin{equation}
\label{1}
\frac{\partial\psi}{\partial t}=(\frac{i}{2m}\Delta-iV(x))\psi
\end{equation}
(and its generalizations that include magnetic fields) can be roughly divided into three
classes. In the approaches of the first class, the
Feynman integral is not supposed to be a genuine integral, but is specified as some
generalized functional on an appropriate space of functions, which can be defined, for
example, as the limit of certain discrete approximations (see e.g. \cite{ET} and more
recent papers \cite{Ich1}, \cite{Ich2}, \cite{Lo}), by means of analytical continuation (see e.g.
\cite{HM}, \cite{JL} and references therein), by extensions of Perceval's identity and by related axiomatic
definitions (see \cite{ABB}, \cite{AKS}, \cite{CW}, \cite{K3}, \cite{SS} and references therein) or by means of the
white noise analysis (see \cite{HKPS}), see also \cite{Za} for the discussion of path integral applied to
the Dirac equation. These approaches still cover only a very restrictive class of potentials, 
for example, singular potentials were considered only by white noise analysis approach but
only in one-dimensional case (see e.g. \cite{AKK} and references therein).

In the approaches of the second class, one tries to define the infinite dimensional Feynman integral as a genuine integral over a bona fide $\sigma$-additive measure on an appropriate
space of trajectories. The first attempt made in \cite{GY} to construct such a measure was
erroneous and led to understanding that there is no direct generalization of Wiener measure
that can give an analog of Feynman-Kac formula for the case of Schr\"odinger operators. A
correct construction of the Feynman integral in terms of the Wiener measure was proposed
in \cite{Do} and was based on the idea of the rotation of the classical trajectories in complex
domains. Namely, changing the variables $x$ to $y$ = $\sqrt ix$ in equation (\ref{1}) leads to the
equation
\[
\frac{\partial\psi}{\partial t}=(-\frac{1}{2m}\Delta-iV(-\sqrt iy))\psi
\]
which is of diffusion type (with possibly complex source) and can be treated by means of
the Feynman-Kac formula and the Wiener measure. Clearly this works only under very
restrictive analytic assumptions on $V$ (see e.g. \cite{H1}, \cite{H2}). However, if one is
interested only in semi classical approximations to the solutions of the Schr\"odinger equation
one can obtain along these lines an approximate path integral representation for even
non-analytic potentials that yields all terms of semi classical expansion (see \cite{BAC}).

Another approach to the construction of the genuine path integral initiated in \cite{M}, 
\cite{MCh} defines it as an expectation with respect to a certain compound Poisson process, or
as an integral over a measure concentrated on piecewise constant paths, see e.g. \cite{Com}, 
\cite{HM}, \cite{K2}, \cite{PQ}, and references therein. Though this method was successfully applied to
different models (see e.g. \cite{GK} for many particle problems, \cite{Se} for simple quantum field
models, \cite{CheQ} for computational aspects and tunneling problems, \cite{Gav} and \cite{KY} for
Dirac equations), the restrictions on interaction forces were always very strong, for example, 
for a usual Schr\"odinger equation, this approach was used only in the case of potentials
which are Fourier transforms of finite measures. However in \cite{K3}, \cite{K4} following this trend, 
a construction was given that covered already essentially more general potentials. To
achieve this, one uses a coordinate representation for the Schr\"odinger equation (and not
the momentum representation as in \cite{MCh}) and also uses an appropriate regularization
of the Schr\"odinger equation.

A more physically motivated regularization to (\ref{1}) (but also technically more difficult
to work with) can be obtained from the theory of continuous quantum measurement (see
\cite{K3}, \cite{K4} for the corresponding results).

In the original papers of Feynman the path integral was defined (heuristically) in
such a way that the solutions to the Schr\"odinger equation were expressed as the integrals
of the function exp$\{iS\}$, where $S$ is the classical action along the paths. It seems that
rigorously the corresponding measure was not constructed even for the case of the heat
equation with sources (notice that in the famous Feynman-Kac formula that gives rigorous
path integral representation for the solutions to the heat equation a part of the action
is actually hidden" inside the Wiener measure).\bigskip

In this article we give a possible definition of the Feynman integral which is related to the first class of approaches above.\bigskip

The mathematically detailed exposition of the result and its proof starts in Section \ref{Seccion 2}, called {\it Algebraic exposition} in a very general and abstract way. There are explanations, motivations, and examples given in this section, of what the objects introduced are good for.

The reader should be willing to go on to Sections \ref{Seccion 3} and \ref{Seccion 4}, which are called {\it Topological exposition} and {\it Analytic exposition}. We define an integral of cylindrical functions'' by means of multiple integrals with respect to finite measures. We define a topology on the space of configurations with which the defined integral has a continuous extension to the set of continuous functions on the space of paths. This extension has a unique representation as a regular and finite Borel measure on the space of paths. In addition, the integral can be defined in terms of a measurable and nonnegative function on the time interval and the space of configurations. Examples hint the reader of the eventual use of the definitions and theorems introduces up to this point.

Sections \ref{Seccion 2} till \ref{Seccion 4} apply to the main results presented in Section \ref{Subseccion 3.3} and Section \ref{Seccion 6}. In Section \ref{Subseccion 3.3}, the Wiener integral is being discussed as an example of this type of integral, and should convince the reader of the power of the notion of Feynman path integral introduced in this paper.

In Section \ref{Seccion 5}, we consider the Schr\"odinger equation for a free particle in a compact space. The fundamental solution of this equation is a distribution. We approximate this distribution by a sequence of smooth functions. We consider the sequence of complex measures corresponding to these functions, as in Section \ref{Seccion 3}. We define a class $S$ of test functions on the space of paths. The class of test functions used here is wide (Cf Section \ref{Seccion 5}), it contains a sufficiently rich space of functions and allows the existence of a Feynman measure. For all test functions, the limit of the integrals of the function with respect to the complex measures exists. We define the Feynman integral of a test function by this limit.
\section{Algebraic exposition}
\label{Seccion 2}
In this section we consider four objects:
\begin{enumerate}
\item an interval $I$, 
\item a real algebra $M$, 
\item a graded real algebra $A$ so that $A_0$ is the set of real numbers, and
\item a collection of $M$-valued functions $f_n$ on the Cartesian product of $I\sp n$ and $ A_n$ such that for all $t$ in $I\sp n$, $f_n$($t$, $\cdot$) is linear.
\end{enumerate}
As with the Wiener measure, we define measures on copies of a variety via the kernel of the Schr\"odinger equation. These measures converge to a measure on the set of curves on the variety.

The interval $I$ is the interval of parameterization of the curves.

The algebra $M$ is the algebra of functions on the space of curves.

The algebra $A_n$ is the algebra of functions on $n$ copies of the variety.

The function $f_n$ is the integral on $n$ copies of the variety. 
\paragraph{Synopsis}
\begin{enumerate}
\item A homomorphism of algebra that extends the action of the functions of the collection, exists (Theorem \ref{Teorema 1}).
\item In Subsection \ref{Subseccion 2.1}, we define a product on the direct sum of the algebras generated by the Cartesian product of $I\sp n$ and $ A_n$ that turns it into an algebra with identity.
\item In Subsection \ref{Subseccion 2.2}, we consider an algebra $M$, an algebra $A$ and a collection of functions $f_1, f_2, ...$ for which the homomorphism of Theorem \ref{Teorema 1} is a homomorphism of algebras.
\item In Subsection \ref{Subseccion 3.1}, we define the homomorphism of Theorem \ref{Teorema 1} by means of multiple integrals with respect to finite measures of bounded and measurable functions. This homomorphism induces a homomorphism in the algebra of functions on the paths on a set.
\end{enumerate}\bigskip

For each natural number $n$, we denote by $M_n$ the free real algebra generated by the Cartesian product of $I\sp n$ and $ A_n$.
\begin{Lema}
There exists a unique real homomorphic extension $h_n$ on $M_n\to M$ of each function of the collection so that the following diagram commutes:
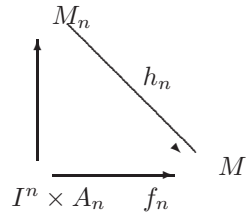
\begin{center}
\setlength{\unitlength}{2500sp}\begingroup\makeatletter\ifx\SetFigFont\undefined\gdef\SetFigFont#1#2#3#4#5{ \reset@font\fontsize{#1}{#2pt} \fontfamily{#3}\fontseries{#4}\fontshape{#5} \selectfont}\fi\endgroup\begin{picture}(1542, 1935)(4759, -4411)
\thinlines
{\put(4801, -3961){\vector( 0, 1){1200}}
}{\put(4951, -4111){\vector( 1, 0){1200}}
}{\put(6226, -3886){\vector( 1, -1){0}}
\multiput(4951, -2611)(6.37500, -6.37500){200}{\makebox(1.6667, 11.6667){\SetFigFont{5}{6}{\rmdefault}{\mddefault}{\updefault}.}}
}\put(4250, -4411){\makebox(0, 0)[lb]{\smash{\SetFigFont{10}{14.4}{\rmdefault}{\mddefault}{\updefault}{$I\sp n\times A_n$}}}}
\put(4650, -2611){\makebox(0, 0)[lb]{\smash{\SetFigFont{10}{14.4}{\rmdefault}{\mddefault}{\updefault}{$M_n$}}}}
\put(6301, -4111){\makebox(0, 0)[lb]{\smash{\SetFigFont{10}{14.4}{\rmdefault}{\mddefault}{\updefault}{$M$}}}}
\put(5551, -4411){\makebox(0, 0)[lb]{\smash{\SetFigFont{10}{14.4}{\rmdefault}{\mddefault}{\updefault}{$f_n$}}}}
\put(5551, -3211){\makebox(0, 0)[lb]{\smash{\SetFigFont{10}{14.4}{\rmdefault}{\mddefault}{\updefault}{$h_n$}}}}
\end{picture}
\end{center}
\end{Lema}
This is a consequence of Proposition 7.49 in \cite{Rotman}.

We denote by
\[
N_n:\hbox{ the sub-algebra generated by the set }
\]
\[
\{(t, a+rb)-(t, a)-r(t, b)\in  M_n\hbox{ so that } r\hbox{ is real}, t\in I\sp n,\hbox{ and }a, b\hbox{ are in } A_n\}, 
\]
\begin{eqnarray*}
M_n/N_n&:&\hbox{the quotient algebra, and}\\
\pi_n&\hbox{on}&M_n\to M_n/N_n, \hbox{ the natural map}.
\end{eqnarray*}
\begin{Lema}
The homomorphism $h_n$ factors through the sub-algebra, i.e., there exists a unique real homomorphism $H_n$ on $M_n$/$N_n\to M$ so that the following diagram commutes:
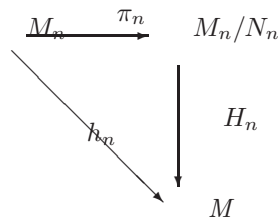
\begin{center}
\setlength{\unitlength}{2500sp}\begingroup\makeatletter\ifx\SetFigFont\undefined\gdef\SetFigFont#1#2#3#4#5{ \reset@font\fontsize{#1}{#2pt} \fontfamily{#3}\fontseries{#4}\fontshape{#5} \selectfont}\fi\endgroup\begin{picture}(1950, 2058)(4501, -4561)
\thinlines
{\put(4801, -2761){\vector( 1, 0){1200}}
}{\put(6301, -3061){\vector( 0, -1){1200}}
}{\put(4651, -2911){\vector( 1, -1){1500}}
}\put(4501, -2761){\makebox(0, 0)[lb]{\smash{\SetFigFont{10}{12.0}{\rmdefault}{\mddefault}{\itdefault}{$M_n$}}}}
\put(6151, -2761){\makebox(0, 0)[lb]{\smash{\SetFigFont{10}{12.0}{\rmdefault}{\mddefault}{\itdefault}{$M_n/N_n$}}}}
\put(6301, -4561){\makebox(0, 0)[lb]{\smash{\SetFigFont{10}{12.0}{\rmdefault}{\mddefault}{\itdefault}{M}}}}
\put(5101, -3811){\makebox(0, 0)[lb]{\smash{\SetFigFont{10}{12.0}{\rmdefault}{\mddefault}{\updefault}{$h_n$}}}}
\put(6451, -3661){\makebox(0, 0)[lb]{\smash{\SetFigFont{10}{12.0}{\rmdefault}{\mddefault}{\updefault}{$H_n$}}}}
\put(5401, -2611){\makebox(0, 0)[lb]{\smash{\SetFigFont{10}{12.0}{\rmdefault}{\mddefault}{\updefault}{$\pi_n$}}}}
\end{picture}
\begin{equation}
\label{2}
\hbox{}
\end{equation}
\end{center}
\end{Lema}
The sub-algebra $N_n$ is in the kernel of the homomorphism $h_n$ ($f_n$($t$, $\cdot$) is linear). Therefore, there exists a unique real homomorphism $H_n$ on $M_n$/$N_n\to M$ so that (\ref{2}) commutes.

We denote by $M_*/N_*$ the direct sum of the algebras $M_n/N_n$.
\begin{Teorema}
\label{Teorema 1}
There exists a homomorphism $H_*$ on $M_*/N_*\to M$ that extends the action of the functions $f_n$.
\end{Teorema}
Consider
\begin{itemize}
\item a set $X$, and
\item a $\sigma$-algebra $\Sigma$ of $X$.
\end{itemize}
\begin{Definicion}
\label{Definicion 2}
For each natural number $n$, 
\begin{enumerate}
\item we denote by $\bigotimes\sp n\Sigma$ the product of $\sigma$-algebras $\Sigma$, 
\item set $A_n$ the real algebra of a collection of bounded and measurable functions on $X\sp n$, 
\item for $t_1$, ..., $t_n$ in $I$,
\begin{itemize}
\item we denote by $\#_{t_1,...,t_n}$ the number of elements of $\{t_1$, ..., $t_n\}$,
\item write $\{t_1$, ..., $t_n\}$ as $\{u_1$, ..., $u_{\#_{t_1,...,t_n}}\}$, and
\item consider the function $f_{t_1,...,t_n}$ on $\{$1, ..., $n\}\to\{$1, ..., $\#_{t_1,...,t_n}\}$ so that
\[
t_i=u_{f_{t_1,...,t_n}(i)}, \qquad i=1, ..., n, 
\]
\end{itemize}
\item\label{Item 5 de Definicion 2}consider a function $m_n$ on the family of the subsets of $I$ of $n$ elements to the set of finite and non negative measures on $\bigotimes\sp n\Sigma$, and
\item set $f_n$ the function on the Cartesian product of $I\sp n$ and $ A_n$ so that
\[
f_n(t_1, ..., t_n; a)
\]
\begin{equation}
\label{5}
=\in t_{X\sp{\#_{t_1,...,t_n}}}a(x_{f_{t_1,...,t_n}(1)}, ..., x_{f_{t_1,...,t_n}(n)})dm_{\#_{t_1,...,t_n}}(\{t_1, ..., t_n\}),
\end{equation}
$t_1$, ..., $t_n$ in $I$, $a$ in $A_n$.
\end{enumerate}
\end{Definicion}
\begin{Ejemplo}\rm
Let $M$ be the real numbers.\\
We will demonstrate that there exists a homomorphism on $M_*/N_*$ to the real numbers that extends the action of the functions $f_n$ (Theorem \ref{Teorema 4}).
\end{Ejemplo}
\subsection{A structure of algebra for the direct sum of the quotient algebras}
\label{Subseccion 2.1}
We define a product on the direct sum of the quotient algebras that turns it into an algebra with identity.
\begin{Definicion}
\label{Definicion 1}
Consider the product on $M_*/N_*$ defined as (($t_1$, $t_2$), $a_1a_2$), $t_i$ in $I\sp{n_i}$ and $a_i$ in $A_{n_i}$ for some $n_i$, $i$=1, 2; so that it distributes the sum.
\end{Definicion}
The product is associative.
\begin{Teorema}
$M_*/N_*$ with the product of Definition \ref{Definicion 1} is a real algebra with identity.
\end{Teorema}
\begin{Ejemplo}\rm
Let
\begin{itemize}
\item$X$ be a locally compact and normal Hausdorff space, and
\item$A_n$ be the real algebra of bounded and continuous functions on $X\sp n$.
\end{itemize}
We will demonstrate that the integral (\ref{5}) has
\begin{itemize}
\item a continuous extension to the set of continuous functions on the curves on $X$ (Theorem \ref{Teorema 5}), and
\item a unique representation as a finite regular Borel measure on the curves on $X$ (Theorem \ref{Teorema 6}).
\end{itemize}
\end{Ejemplo}
\subsection{The homomorphism of Theorem \ref{Teorema 1} as homomorphism of algebras}
\label{Subseccion 2.2}
We consider an algebra $M$, an algebra $A$ and a collection of functions $f_1, f_2, ...$ for which the homomorphism of Theorem \ref{Teorema 1} is a homomorphism of algebras.\bigskip

Consider
\begin{itemize}
\item a nonempty set $X$, and
\item the indexed family $\{X_t$\hbox{ so that }$t\in I\}$ so that
\[
 X_t=X,\quad t\in  I.
\]
\end{itemize}
We denote by
\begin{eqnarray*}
\prod_tX_t&:&\hbox{the product of the }X_t, \hbox{ and}\\
p_t&\hbox{on}&\prod_{u}X_{u}\to X_t, \hbox{ the projections}.
\end{eqnarray*}

Set $M$ the real algebra of functions on $\prod_tX_t$.

For each natural number $n$, 
\begin{itemize}
\item set $A_n$ the real algebra of a collection of functions on $X\sp n$, and
\item consider the function $f_n$ on the Cartesian product of $I\sp n$ and $ A_n\to M$ so that
\begin{equation}
\label{4}
f_n(t_1, ..., t_n; a)[\pi]=a(p_{t_1}(\pi), ..., p_{t_n}(\pi)),\quad\pi\in \prod_tX_t.
\end{equation}
\end{itemize}
By (\ref{4}), it is easy to prove that the function $f_n$($t$, $\cdot$) is linear, $t$ in $I\sp n$.
\begin{Lema}
\label{Lema 4}
The homomorphism of Theorem \ref{Teorema 1} is a homomorphism of real algebras, i.e.,
\[
H_{n+o}(t, u; ab)=H_{n}(t, a)H_{o}(u, b).
\]
\end{Lema}
Write $t$ as ($t_1$, ..., $t_n$) and $u$ as ($u_1$, ..., $u_o$).
\begin{eqnarray*}
H_{n+o}(t, u; ab)[\pi]&=&(ab)(p_{t_1}(\pi), ..., p_{t_n}(\pi), p_{u_1}(\pi), ..., p_{u_o}(\pi))\\
&=&a(p_{t_1}(\pi), ..., p_{t_n}(\pi))b(p_{u_1}(\pi), ..., p_{u_o}(\pi))\\
&=&H_{n}(t, a)[\pi]H_{o}(u, b)[\pi].
\end{eqnarray*}
\begin{Teorema}
\label{Teorema 3}
The homomorphism $H_*$ on $M_*/N_*\to M$ is a homomorphism of real algebras.
\end{Teorema}
This is a consequence of Definition \ref{Definicion 1} and Lemma \ref{Lema 4}.
\begin{Ejemplo}\rm
Let $X$ be a complete Riemann manifold.\\
We will define a measure on the space of curves on $X$ in terms of a measurable and nonnegative function on the interval of parameterization and the manifold. An example of this type of measure is the Wiener measure.
\end{Ejemplo}
\subsection{The homomorphism induced in the algebra of functions on $\prod_tX_t$}
\label{Subseccion 3.1}
\begin{Definicion}
\label{Definicion 3}
Consider the homomorphism $i_*$ on $H_*M_*$/$N_*$ to the real numbers so that the following diagram commutes:
\begin{center}
\setlength{\unitlength}{2500sp}\begingroup\makeatletter\ifx\SetFigFont\undefined\gdef\SetFigFont#1#2#3#4#5{ \reset@font\fontsize{#1}{#2pt} \fontfamily{#3}\fontseries{#4}\fontshape{#5} \selectfont}\fi\endgroup\begin{picture}(1800, 2070)(4501, -5761)
\thinlines
{\put(4651, -4111){\vector( 0, -1){1200}}
}{\put(4801, -5461){\vector( 1, 0){1200}}
}{\put(6151, -4111){\vector( 0, -1){1200}}
}{\put(4801, -3961){\vector( 1, 0){1200}}
}\put(4000, -3961){\makebox(0, 0)[lb]{\smash{\SetFigFont{10}{12.0}{\rmdefault}{\mddefault}{\itdefault}{M$_*$/N$_*$}}}}
\put(3750, -5611){\makebox(0, 0)[lb]{\smash{\SetFigFont{10}{12.0}{\rmdefault}{\mddefault}{\itdefault}{H$_*$M$_*$/N$_*$}}}}
\put(6151, -3961){\makebox(0, 0)[lb]{\smash{\SetFigFont{10}{12.0}{\rmdefault}{\mddefault}{\itdefault}{\mbox{I\negthinspace R}}}}}
\put(6151, -5611){\makebox(0, 0)[lb]{\smash{\SetFigFont{10}{12.0}{\rmdefault}{\mddefault}{\itdefault}{\mbox{I\negthinspace R}}}}}
\put(4250, -4711){\makebox(0, 0)[lb]{\smash{\SetFigFont{10}{12.0}{\rmdefault}{\mddefault}{\itdefault}{H$_*$}}}}
\put(5251, -3811){\makebox(0, 0)[lb]{\smash{\SetFigFont{10}{12.0}{\rmdefault}{\mddefault}{\itdefault}{h$_*$}}}}
\put(5251, -5761){\makebox(0, 0)[lb]{\smash{\SetFigFont{10}{12.0}{\rmdefault}{\mddefault}{\itdefault}{i$_*$}
}}}
\put(6301, -4711){\makebox(0, 0)[lb]{\smash{\SetFigFont{10}{12.0}{\rmdefault}{\mddefault}{\rmdefault}{=}
}}}
\end{picture}
\begin{equation}
\label{7}
\hbox{}
\end{equation}
\end{center}
\end{Definicion}
\begin{Teorema}
\label{Teorema 4}
There exists a homomorphism $h_*$ on $M_*/N_*$ to the real numbers that extends the action of the functions $f_n$.
\end{Teorema}
By (\ref{5}), it is easy to prove that the function $f_n$($t$, $\cdot$) is linear, $t$ in $I\sp n$. Then, the theorem is a consequence of Theorem \ref{Teorema 1}, where in this case $M$ is the set of real numbers.
\section{Topological exposition}
\label{Seccion 3}
We define a topology on the set $X$ so that the induced homomorphism has a continuous extension to the set of continuous functions on $\prod_tX_t$.

Suppose that
\begin{itemize}
\item$X$ is a locally compact and normal Hausdorff space, and
\item$\Sigma$ is so that for every natural number $n$, the continuous functions on $X\sp n$ are measurable.
\end{itemize}

Set $A_n$ the real algebra of bounded and continuous functions on $X\sp n$.

Consider the product topology on $\prod_tX_t$.

We denote by $C$($\prod_tX_t$) the set of continuous functions on $\prod_tX_t$ with the uniform metric.\begin{Lema}
\label{Lema 6}
The composition of a continuous function $f$ on the Cartesian product of copies of $X$ with projections $p_t$, ..., $p_u$; i. e., $f$($p_t$, ..., $p_u$), is continuous on $\prod_tX_t$.
\end{Lema}
Each projection $p_t$ is continuous (Theorem 3.7 in \cite{Sieradski}). Therefore, the function $f$($p_t$, ..., $p_u$) is continuous.
\begin{Lema}
\label{Lema 7}
Given two points $\pi$, $\varpi$ of the product $\prod_tX_t$, there exist (i) an element $t$ in $I$, and (ii) a bounded and continuous function $f$ on $X$, such that $f$ separates the images of the points under the projection $p_t$:
\[
f(p_t(\pi))\neq f(p_t(\varpi)).
\]
\end{Lema}
There exists an element $t\in  I$ so that the corresponding projection separates the points:
\[
p_t(\pi)\neq p_t(\varpi).
\]
By the Theorem of Urysohn (\cite{Sieradski}, page 238), there exists a continuous function $f$ on $X$ to [0, 1], so that it separates the images of the points under the projection.
\begin{Definicion}
If $X$ is not compact, then by Alexandroff's Theorem (\cite{Sieradski}, page 217), $X$ has a Hausdorff one-point compactification, which we denote by $X$ also.
\end{Definicion}
\begin{Lema}
\label{Lema 8}
$\prod_tX_t$ is compact.
\end{Lema}
This is a consequence of Tychonoff's Theorem (\cite{Sieradski}, page 205).
\begin{Lema}
\label{Lema 9}
The image of the homomorphism of Theorem \ref{Teorema 3}, $H_*M_*$/N$_*$, is a dense subspace of the set of continuous functions $C$($\prod_tX_t$).
\end{Lema}
The image of the homomorphism $H_*M_*$/N$_*$ is contained in the set of continuous functions $C$($\prod_tX_t$) (Lemma \ref{Lema 6}). By the Theorem of Stone and Weierstrass (\cite{Hewitt}, page 95), $H_*M_*$/N$_*$ is dense in $C$($\prod_tX_t$) (also see Lemmas \ref{Lema 7} and \ref{Lema 8}).
\begin{Teorema}
\label{Teorema 5}
The homomorphism of Definition \ref{Definicion 3}, $i_*$, has a linear and continuous extension $F$ to the set of continuous functions $C$($\prod_tX_t$).
\end{Teorema}
This is a consequence of Lemma \ref{Lema 9} and the inequality
\[
|f_n(t_1, ..., t_n; a)|
\]
\[
\le\in t_{X\sp{\#_{t_1,...,t_n}}}\sup|a|dm_{\#_{t_1,...,t_n}}(\{t_1,...,t_n\}),\quad t_1, ..., t_n\in  I; a\in A_n.
\]
\section{Analytical exposition}
\label{Seccion 4}
The extension of Theorem \ref{Teorema 5} has a unique representation as a finite regular Borel measure on $\prod_tX_t$.

In Subsection \ref{Subseccion 3.3}, we define a measure on the space of curves on a complete Riemann manifold in terms of a measurable and nonnegative function on the interval of parameterization and the manifold. An example of this type of measure is the Wiener measure.
\begin{Lema}
\label{Lema 10}
The linear functional $F$ is nonnegative: $Ff\ge$0, $f$ a continuous and positive function on $\prod_tX_t$.
\end{Lema}
The linear functional $F$ is continuous (Theorem \ref{Teorema 5}): for all $\epsilon>$0, exists $\delta>$0 such that
\begin{equation}
\label{9}
\sup|f-c|<\delta\hbox{ implies } Ff>Fc-\epsilon,\quad c\in  C\left(\prod_tX_t\right).
\end{equation}
The minimum of $f$ is greater than Zero (Lemma \ref{Lema 8}). The image of the homomorphism $H_*$ of Theorem \ref{Teorema 3} is dense in $C$($\prod_tX_t$) (Lemma \ref{Lema 9}): there exists $m$ in $M_*/N_*$ such that
\begin{equation}
\label{10}
|f-H_*m|(\pi)<\min\{\in f f, \delta\},\quad\pi\in \prod_tX_t,
\end{equation}
in particular, 
\begin{equation}
\label{8}
(H_*m)(\pi)>f(\pi)-\in f f.
\end{equation}
By (\ref{10}) and (\ref{9}), 
\begin{eqnarray*}
Ff&>&FH_*m-\epsilon\\
&=&i_*H_*m-\epsilon\quad(\hbox{by Theorem }(\ref{Teorema 5}))\\
&=&h_*m-\epsilon\quad(\hbox{by }(\ref{7}))\\
&>&-\epsilon, 
\end{eqnarray*}
by (\ref{8}). Therefore, $Ff\ge$0.
\begin{Lema}
\label{Lema 11}
Consider a continuous function $f$ on $\prod_tX_t$ to [0, $\in fty$). Then, $Ff\ge$0.
\end{Lema}
Notice that
\[
\max\{\delta, f(\pi)\}-f(\pi)\le\delta,\quad\delta>0, \pi\in \prod_tX_t,
\]
and since the linear functional $F$ is continuous (Theorem \ref{Teorema 5}): for all $\epsilon>$0, exists $\delta>$0 such that
\begin{eqnarray*}
 Ff&\ge&F\max\{\delta, f\}-\epsilon\\
&\ge&-\epsilon, 
\end{eqnarray*}
by Lemma \ref{Lema 10}, where in this case $f$"=max$\{\delta$, $f\}$. Therefore, $Ff\ge$0.
\begin{Teorema}
\label{Teorema 6}
The linear functional $F$ has a unique representation as a finite regular Borel measure $m$ on $\prod_tX_t$:
\[
Fc=\in t_{\prod_tX_t}c\, dm,\quad c\in  C\left(\prod_tX_t\right).
\]
\end{Teorema}
This is a consequence of Theorem \ref{Teorema 5}, Lemma \ref{Lema 11} and the Riez Representation Theorem (Theorem 22.8 in \cite{Nielsen}).
\section{Measures in the space of curves on a complete Riemann manifold}
\label{Subseccion 3.3}
We define a measure on the space of curves on a complete Riemann manifold in terms of a measurable and nonnegative function on the interval of parameterization and the manifold. An example of this type of measure is the Wiener measure.\bigskip

Set $X$ a complete Riemann manifold.

We denote by $dx$ the Riemann volume measure.
\begin{Definicion}
\label{Definicion 5}
Consider
\begin{itemize}
\item an element $x_0\in X$, and
\item a measurable function $f$ on the Cartesian product of $I^2$ and $ X^2$, to [0, $\infty$) so that for $t_0$, ..., $t_n$ in $I$, $\prod_{i=1}\sp nf$($t_{i-1}$, $t_i$, $x_{i-1}$, $x_i$)$dx_i$ is a finite measure on $X\sp n$.
\end{itemize}
Order $t_0$, ..., $t_n$ as $t_0'<\cdots<t_{\#_{t_0,...,t_n}}'$, and in Item \ref{Item 5 de Definicion 2} of Definition \ref{Definicion 2}, set
\[
dm_n(\{t_0, ..., t_n\})=\prod_{i=1}\sp{\#_{t_0,...,t_n}}f(t_{i-1}', t_i', x_{i-1}, x_i)dx_i.
\]
For this $f$, we define the regular and finite Borel measure on $\prod_tX_t$ of Theorem \ref{Teorema 6}.
\end{Definicion}
\begin{Ejemplo}[The Wiener measure]\rm
The Wiener measure involves a reference to the existence of a unique fundamental solution of the heat equation on a complete manifold. This affirmation is considerably older than the one we quote here \cite{Dodziuk}.

We denote by
\begin{eqnarray*}
g&:&\hbox{the metric tensor}, \\
\hbox{Ric}&:&\hbox{the Ricci tensor, and}\\
\Delta&:&\hbox{the Laplacian acting on }C\sp\infty(X).
\end{eqnarray*}
Suppose that there exists $c>$0 so that Ric$\ge$-$cg$. By Theorem 4.2 in \cite{Dodziuk}, a unique fundamental solution $f$($t$, $x$, $y$) of the equation
\[
\frac{\partial u}{\partial t}=\Delta u, 
\]
exists. The measure defined from the function $f$ is the Wiener measure.
\end{Ejemplo}
\subsection{The Feynman integral of a test function}
\label{Seccion 5}
\begin{Definicion}
\label{Definicion 6}
Consider a function $f$ on $X$ to the set of the complex numbers, so that $ \pm$max$ \{$0, $ \pm \Im f \} $ and $ \pm$max$ \{$0, $ \pm \Re f \} $ satisfy the hypothesis in Definition \ref{Definicion 5}. We will denote by $m_ {\pm \Im f} $ and $m_ {\pm \Re f} $ the regular and finite Borel measures corresponding to $ \pm$max$ \{$0, $ \pm \Im f \} $ and $ \pm$max$ \{$0, $ \pm \Re f \} $ according to Definition \ref{Definicion 5}, respectively. We define the complex measure $m_{\Re f}$-$m_{-\Re f}$+$i$($m_{\Im f}$-$m_{-\Im f}$).
\end{Definicion}

Consider a differential operator $O$ that acts on functions on $X$. The fundamental solution of the equation
\[
-i\frac{\partial u}{\partial t}=O u, 
\]
is a distribution. We approximate this distribution by a sequence $f_1$, $f_2$, ... of smooth functions on the Cartesian product of $I$ and $X^2$, i.e.; for a test function $\phi$ on the Cartesian product of $I$ and $X$,
\[
\lim_{k\to\infty}\left(i\frac\partial{\partial t}+O\right)\int_I\int_Xf_k(t, u, x, y)\phi(u, y)dydu=\phi(t, x).
\]
Such sequence satisfies the hypothesis in Definition \ref{Definicion 6}. We consider the sequence of complex measures corresponding to this sequence of functions, according to Definition \ref{Definicion 6}.
\begin{Definicion}
We consider the set of test functions on the space of curves of the manifold for which the limit of the integrals of a function with respect to the complex measures exists, i.e.; for
\begin{itemize}
\item a natural number $n$,
\item a partition $t_0<t_1<$ ... $<t_n$ of $I$, and
\item a function $\phi$ on $X^n$,
\[
\lim_{k\to\infty}\int_{X^n}\phi(x_1, ..., x_n)\prod_{i=1}^nf_k(t_{i-1}, t_i, x_{i-1}, x_i)dx_i
\]
exists.
\end{itemize}
We define the integral of Feynman of $\phi$ by this limit.
\end{Definicion}
 Important for the usefulness of any approach to Feynman integrals is the class of potentials we are able to handle.
\section{The Feynman integrand for a new class of unbounded potentials}
\label{Seccion 6}
We include some examples concerning the Feynman propagator for special potentials. These examples show the possibility of applying author's Feynman integral as a tool for constructing dynamics with a potential.

We begin by discussing a free motion on a manifold. Even for this restricted class of systems the existence of a measure on paths is interesting. We will consider the equation of eigen-values of the dynamic Schr\"odinger operator in polar coordinates without potential, i.e.; consider
\begin{itemize}
\item three numbers $n$, $ \lambda$ and $ \nu$, and
\item the equation
\begin{equation}
\label{3}
-i \frac{\partial f}{\partial t}-{r^{1-n}}\frac\partial{\partial r} \left({r^{-1+n}}\frac{\partial f}{\partial r}\right)+\frac{\nu}{{r^2}} f=\lambda f.
\end{equation}
\end{itemize}
Set
\[
o={\sqrt{{{\left(-1+\frac{n}{2}\right)}^2}+\nu }}.
\]
We will denote by $B_o$ the function of Bessel of order $o$. We will suggest an integral transformation in terms of the functions of Bessel as a solution, i.e.; consider a solution of (\ref{3}) of the form
\begin{equation}
\label{11}
{r^{1-\frac{n}{2}}}\int_0^\infty {B_o}(k r)F(k, t)dk.
\end{equation}
The functions of Bessel satisfy 
\begin{eqnarray}
\label{12}
2B_o'&=&B_{o-1}-B_{1+o},\\
\label{6}
\frac{2o}xB_o(x)&=&B_{o-1}(x)+B_{1+o}(x).
\end{eqnarray}
We replace (\ref{11}) in (\ref{3}). We use (\ref{12}) and (\ref{6}).

(\ref{3}) becomes
\[
{r^{1-\frac{n}{2}}}\int_0^\infty {B_o}(k r) \left(\left({k^2}-\lambda \right) F(k,t)-{i} {\frac\partial{\partial t}} F(k,t)\right)=0.
\]
Therefore, the equation of eigen-values is simplified to an ordinary equation in the time variable
\begin{equation}
\label{13}
\left({k^2}-\lambda \right) F(k,t)-{i} {\frac\partial{\partial t}} F(k,t)=0.
\end{equation}
{exp}$\left(-{i} t \left({k^2}-\lambda \right)\right)$ solves (\ref{13}).

We calculate
\[
{r^{1-\frac{n}{2}}}\int_0^\infty \hbox{exp}\left(-{i} t \left({k^2}-\lambda \right)\right){B_o}(k r)dk
\]
\begin{equation}
\label{14}
=\frac{e^{{i} \left(t \lambda +\frac{1}{4} \left(\frac{{r^2}}{2 t}-(1+o) \pi \hbox{\small sign}t\right)\right)} {\sqrt{\pi }} {r^{1-\frac{n}{2}}} {B_{\frac{o}{2}}}\left(\frac{{r^2}}{8 |t|}\right)}{2 {\sqrt{|t|}}}.
\end{equation}
We denote by $p$($r$, $t$, $\lambda$) the right hand side of (\ref{14}).

We calculate the propagator of the Schr\"odinger equation with the solution of the equation of eigen-values.
\[
\int_{-\infty}^\infty\frac{p(r, t, \lambda)p(s, u, \lambda)}\lambda d\lambda
\]
\[
=\frac{e^{\frac{i}4\left(\frac{{r^2}}{2t}-\frac{{s^2}}{2u}+(1+o) \pi (\hbox{\small sign} u-\hbox{\small sign} t)\right)}{{\pi }^2} {(rs)^{1-\frac{n}{2}}}  {B_{\frac{o}{2}}}\left(\frac{{r^2}}{8 |t|}\right) {B_{\frac{o}{2}}}\left(\frac{{s^2}}{8 |u|}\right) \hbox{sign}(t-u)}{4 {\sqrt{|t|}} {\sqrt{|u|}}}.
\]

We now construct the Feynman integrand for a new class of potentials and calculate the propagators. Any new method has to provide a decent class of $V$.

Consider a function $V$ on (0, $ \infty$) so that for $t \neq$0, $ \int_0^ \infty {r^ {2-n}} {B_{\frac o2} ^2} (\frac {{r^2}} {8 |t|}) V$($r$)$dr$ is finite. This integral is finite if, for example, exists an exponent $e>$ -1 so that
\[
V(r)=r^e
\]
near Zero. Consider the function 
\[
q(r, t)=\frac{\hbox{exp}\left(\frac{i}4 \left(\frac{{r^2}}{2 t}-(1+o) \pi \hbox{sign} t\right)\right)\pi {r^{1-\frac{n}{2}}} {B_{\frac{o}{2}}}\left(\frac{{r^2}}{8 |t|}\right)}{2 {\sqrt{|t|}}}.
\]

The propagator of the equation
\[
-i \frac{\partial f}{\partial t}-{r^{1-n}}\frac\partial{\partial r} \left({r^{-1+n}}\frac{\partial f}{\partial r}\right)+\frac{\nu}{{r^2}} f=0
\]
has a perturbation series which is uniformly absolutely convergent in the time for every compact set in the variables $r$, $s$
\[
f_k(r, s; t, u)=q(r, t)\bar q(s, u)\sum_{i=0}^k\prod_{j=1}^i\int_0^\infty|q(x, t_j)|^2V(x)dx,
\]
$t<t_1<$ ... $<t_k<u$.

Since $ \int_0^ \infty {r^ {1 \frac n2}} {B_{\frac o2}} (\frac {{r^2}} {8 |t|}) dr$ and $ \int_0^ \infty {r^ {2-n}} {B_{\frac o2} ^2} (\frac {{r^2}} {8 |t|}) dr$ are finite, Definition \ref{Definicion 6} applies.

\end{document}